\documentclass[conference]{IEEEtran}
\IEEEoverridecommandlockouts
 \usepackage{geometry}
\usepackage{enumitem} 
\usepackage{cite}
\usepackage{amssymb}
\usepackage{amsmath}
\usepackage{algorithmic}
\usepackage{graphicx}
\usepackage{textcomp}
\usepackage{xcolor}
\usepackage{multicol}
\usepackage{multirow}
\usepackage{booktabs}
\usepackage{ntheorem}
\usepackage{diagbox}
\usepackage{makecell}

\def\BibTeX{{\rm B\kern-.05em{\sc i\kern-.025em b}\kern-.08em
		T\kern-.1667em\lower.7ex\hbox{E}\kern-.125emX}}
\newtheorem{remark}{Remark}

\theoremstyle{nonumberplain}

\begin{document}
	
	\title{\huge RIS-Assisted MIMO Communication Systems: Model-based versus Autoencoder Approaches \vspace{-0.4cm}\\
		
	}
	
	\author{\IEEEauthorblockN{Ha An Le$^{\ast}$, Trinh Van Chien$^{\nu}$, Van Duc Nguyen$^{\dagger}$, and Wan Choi$^\ast$ }
		\IEEEauthorblockA{$^{\ast}$Department of Electrical and Computer Engineering, Seoul National University, Seoul, Korea\\
			$^\nu$School of Information and Communications Technology (SoICT), Hanoi University of Science and Technology, Vietnam
			\\
			$^{\dagger}$School of Electrical and Electronic Engineering, Hanoi University of Science and Technology, Hanoi, Vietnam
	\\
			Emails: 25251225@snu.ac.kr,
			chientv@soict.hust.edu.vn, duc.nguyenvan1@hust.edu.vn, wanchoi@snu.ac.kr
		\vspace{-0.9cm}}
\thanks{This work was supported by Institute of Information \& communications Technology Planning \& evaluation (IITP) grant funded by the Korea government (MSIT) (No. 2018-0-00809, Development on the disruptive technologies for beyond 5G mobile communications employing new resources).}
			\vspace*{-0.15cm}
	}

	\maketitle
	
	\begin{abstract}		 
	This paper considers reconfigurable intelligent surface (RIS)-assisted point-to-point multiple-input multiple-output (MIMO) communication systems, where a transmitter communicates with a receiver through an RIS. Based on the main target of reducing the bit error rate (BER) and therefore enhancing the communication reliability, we study different model-based and data-driven (autoencoder) approaches. In particular, we consider a model-based approach that optimizes both active and passive optimization variables. We further propose a novel end-to-end data-driven framework, which leverages the recent advances in machine learning. The neural networks presented for conventional signal processing modules are jointly trained with the channel effects to minimize the bit error detection. Numerical results demonstrate that the proposed data-driven approach can learn to encode the transmitted signal via different channel realizations dynamically. In addition, the data-driven approach not only offers a significant gain in the BER performance compared to the other state-of-the-art benchmarks but also guarantees the performance when perfect channel information is unavailable.
	\end{abstract}
	\begin{IEEEkeywords}
    Reconfigurable Intelligent Surface, Multiple-Input Multiple-Output, Autoencoder, Bit Error Rate.
\end{IEEEkeywords}
	\vspace*{-0.2cm}
	\section{Introduction}
	\vspace*{-0.1cm}
	
	Reconfigurable intelligent surface (RIS) has been recently emerged as a cost-effective paradigm that can tackle the complexity issues \cite{Qingqing2021IRS}. Each RIS device  comprises a  number of  reflecting elements that can be controlled to manipulate the incoming signals in the desired manner. With a proper phase shift design, the reflected signal can be added constructively  to enhance the signal strength  \cite{Qingqing2021IRS}.  The authors in \cite{QingQing2019IRS} proposed a joint active and passive beamforming design in MISO multi-user system to minimize the total transmit power consumption. In \cite{Huang-RIS2019}, an alternating maximization approach has been applied to jointly optimize precoding matrix and RIS phase-shifts in order to boost energy efficiency of the system. Consider a RIS-assisted point-to-point MIMO system, the authors in \cite{ZhangRISMIMO2020} investigated the optimization of RIS phase-shift in OFDM MIMO system to enhance to capacity of the cascaded channel. However, in \cite{ZhangRISMIMO2020}, a precoding matrix at the transmitter has not been considered. In \cite{Boyu2020}, a low complexity design of precoding matrix and phase-shift of RIS has been proposed to maximize the spectral efficiency of a point-to-point MIMO system. Related to the communication reliability, there are only a few works. In \cite{Aymen2021}, an RIS-based Vertical Bell Labs layered space-time (VBLAST) system has been proposed to enhance the bit error rate (BER) performance of the MIMO system. These previous works suffer from high computational complexity since the algorithms are implemented in iterative manners. The local solutions from the model-based approaches might be much worse than the global optimum,  motivating for a better design.
	
	Various applications of machine learning in wireless communications such as resource management, channel estimation, and signal detection \cite{Chien2020PowerControl} have been proposed to address intractable non-convex optimization problems and high-complexity issues. Related to the RIS system, the idea of applying machine learning in designing RIS phase-shift has been widely studied. In \cite{Huang2019-RISDNN}, the authors proposed a deep neural network (DNN) to learn the optimal RIS phase-shift from the users'
	positions in order to maximize the through put of the multi-user MISO system. In \cite{Song2021-RISCNN}, a framework which is comprised of two convolutional neural networks (CNNs) is utilized to jointly optimize the RIS phase-shift and precoding matrix in an unsupervised fashion with the goal of maximizing the sum rate of all users. With the {same optimization objective with \cite{Song2021-RISCNN}}, in \cite{Huang2020-RLRIS}, a reinforcement learning-based framework is proposed to predict the optimal precoding matrix and RIS phase-shift with the given channel realizations. Although different machine learning approaches have been successfully applied in RIS systems, there is very limit work that focuses on improving {signal detection performance} and reliability of the system. 
	In \cite{Tugba2021AE-RIS}, the authors considered a jointly transmitter, receiver and RIS phase-shift design. In this framework, an autoencoder approach is proposed to enhance the BER performance of RIS-assited SISO system. Furthermore, the authors in \cite{Jiang2022-EndToEndRIS} proposed a autoencoder approach-based design for RIS-assited MIMO systems to improve data detection performance. However, the works in \cite{Tugba2021AE-RIS} and \cite{Jiang2022-EndToEndRIS} are limited since the autoencoders are trained based on one deterministic channel realization. This not only makes the autoencoder can work only for one specific channel but also does not really reflex the practical aspects of physical radio channels.\footnote{The previous works \cite{Tugba2021AE-RIS,Jiang2022-EndToEndRIS} assumed sufficiently large coherent time, so their autoencoder architectures may not enable them to cope with the rapid changes of propagation channels in practice.} 
	Motivated by this drawback, we propose an autoencoder approach that can encode data through different channel realizations to enhance the BER.

	\textit{Paper contributions}: In this paper, for the model-based approach, we first study the two different MIMO communication system models that optimize radio resource and smartly control the propagation environments with the BER minimization as the utility metric. For the data-driven approach, we propose an autoencoder design for RIS-assisted MIMO communication systems, which consider the practical conditions of fading channels, and therefore close to the real applications than previous works. We jointly design the RIS phase-shift and the transceiver in order to reduce the BER at the receiver. Following the end-to-end framework, the transmitter, receiver, and the RIS device are modeled by individual deep neural networks with a unique loss function that minimizes the bit detection error on average. Our proposed data-driven approach establishes a unified framework to learn the entire system in order to enhance the communication reliability. Numerical results demonstrates the benefits of the data-driven approach that  provides much better BER performance than the other model-based approaches. Moreover, we numerically show that the proposed framework can learn to encoder data  that are robust to the fluctuation of propagation channels, and therefore retains a good BER performance under imperfect channel state information (CSI).
	
	\textit{Notation}: The upper and lower bold letters are utilized to denote the matrices and vectors. The superscript $(\cdot)^T$ and $(\cdot)^H$ are the regular and Hermitian transpose. $\mathbf{I}_N$ denotes an identity matrix of size $N \times N$ and $\mathrm{arg}(\cdot)$  is the argument of a complex number. $\| \cdot \|$ and $\| \cdot \|_F$ denote the Euclidean and Frobenius norm. The expectation of a random variable is denoted by $\mathbb{E}\{\cdot\}$, while $\mathcal{CN}(\cdot, \cdot)$ is a circularly symmetric Gaussian distribution. Finally, $\mathcal{O}(\cdot)$ is the big-$\mathcal{O}$ notation.
	\begin{figure}[t]
		\centering
		\includegraphics[trim=0.1cm 0.5cm 0cm 0.1cm, clip=true, width=2.7in]{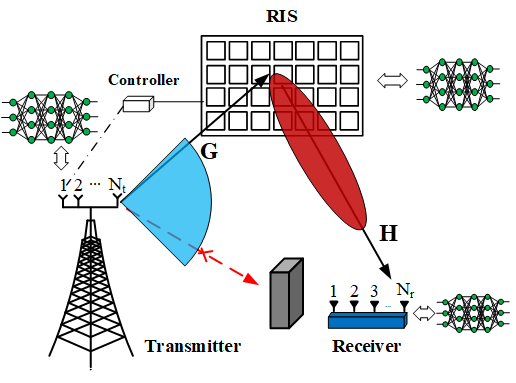} \vspace*{-0.25cm}
		\caption{The considered  RIS-assisted MIMO communication  system model where the transceiver and the RIS are replaced by neural networks.}
		\label{Fig:SystemModel}
		\vspace*{-0.4cm}
	\end{figure}
	\vspace*{-0.2cm}
	\section{MODEL-BASED OPTIMIZATION APPROACH} \label{Sys-Model}
	\vspace*{-0.2cm}
	We consider a point-to-point MIMO system where a transmitter equipped with $N_t$ antennas transmits $N_s$ data streams to the receiver. The receiver has  $N_r$ antennas  to enhance the received signal strength. Furthermore, the system performance is boosted by the support of an RIS comprising $K$ passive reflecting elements  as illustrated in Fig.~\ref{Fig:SystemModel}. The reflection matrix $\pmb{\Theta} \in \mathbb{C}^{K \times K}$ is formulated as
	  	\vspace*{-0.1cm}
		\begin{equation}
			\pmb{\Theta} =  \mathrm{diag}(\beta_1 e^{j\theta_1},\dots, \beta_K e^{j\theta_K}),
			\vspace*{-0.1cm}
		\end{equation}
		where $0 \leq \beta_k \leq 1$ and $-\pi \leq \theta_k \leq \pi$ are the magnitude and phase produced by the $k$-th reflecting element. As a popular assumption because of the recent advances towards lossless metasurfaces, we assume a unit signal reflection, i.e., $\beta_k = 1, \forall k$. In this paper, we assume that the direct link between the transmitter and the receiver is blocked due to large obstacles. The data is sent from the transmitter to the receiver through the RIS by the indirect link with the cascaded channels. 
		 The signal $\mathbf{s}$ is assumed to be $M$-QAM (quadrature amplitude modulation) with $\mathbb{E}\{ \mathbf{s} \mathbf{s}^H \} = \mathbf{I}_{N_t}$. 
		The modulated data message is  precoded by a linear precoder $\mathbf{F} \in \mathbb{C}^{N_t \times N_s}$ with $\|\mathbf{F}\|^2_F = N_s$, and then passed through to the receiver via the RIS. The 
		received signal, $\mathbf{y} \in \mathbb{C}^{N_r}$, at the receiver is 
		\vspace*{-0.1cm}
		\begin{equation} \label{eq:ReceivedSig}
			\mathbf{y} = \sqrt{P/N_s}  \mathbf{H}^H\pmb{\Theta} \mathbf{G}\mathbf{F}\mathbf{s} + \mathbf{n},
		\vspace*{-0.1cm}
		\end{equation}
		where $P$ is the total transmitted power. The channel between the transmitter and the RIS is denoted by  $\mathbf{G} \in \mathbb{C}^{K \times N_t}$, while $\mathbf{H} \in \mathbb{C}^{K \times N_r}$ is the channel between the RIS and the receiver.   Additionally, $\mathbf{n} \sim \mathcal{CN}(\mathbf{0},\sigma^2\mathbf{I}_{N_r})$ denotes additive white Gaussian noise (AWGN). {We assume that all the channels are known to the transmitter with the help of feedback or channel reciprocity.} From \eqref{eq:ReceivedSig}, the signal is then decoded as
		\vspace*{-0.1cm}
		\begin{equation} \label{eq:shat}
			\hat{\mathbf{s}} =  \mathbf{Z}\mathbf{y} = \sqrt{P/N_s}  \mathbf{Z} \mathbf{H}^H\pmb{\Theta} \mathbf{G}\mathbf{F}\mathbf{s} + \mathbf{Z} \mathbf{n},
		\vspace*{-0.1cm}
		\end{equation}
		where $\mathbf{Z} \in \mathbb{C}^{N_s \times N_r}$ represents the equalizer matrix. The active beamforming matrices, i.e., the precoding matrix $\mathbf{F}$ and the equalizer matrix $\mathbf{Z}$, together with  the passive reflection matrix $\pmb{\Phi}$ can be optimized for a given utility metric and practical constraints. However, the strong coupling among these optimization variables leads to a nontrivial procedure to solve the resource allocation problems optimally. In this paper, we separately solve subproblems $\pmb{\Theta}$ and $\mathbf{F}$ to obtain a good solution in polynomial time by, first, maximizing the channel capacity defined for \eqref{eq:ReceivedSig}. Then, based on \eqref{eq:shat}, the channel impairment is compensated by a proper selection of the equalizer matrix $\mathbf{Z}$.
	\subsubsection{Design of the reflection matrix $\pmb{\Theta}$} Let us define $\mathbf{f}_0(\pmb{\Theta}, \mathbf{Z}, \mathbf{F}) = (P/N_s)\mathbf{Z}\mathbf{H}^H \pmb{\Theta} \mathbf{G} \mathbf{F}\mathbf{F}^H \mathbf{G}^H \pmb{\Theta}^H  \mathbf{H} \mathbf{Z}^H$. Following the same methodology as \cite[Proposition 1]{Boyu2020} with the given optimal solution to the precoding matrix $\mathbf{F}$ and the equalizer $\mathbf{Z}$, it holds that  $\log_2\det|\mathbf{I}_{N_r}+ \mathbf{f}_0(\pmb{\Theta}, \mathbf{Z}, \mathbf{F}) | \geq \log_2( 1 + (P/N_s)\mathrm{tr}(  \mathbf{H}^H\pmb{\Theta} \mathbf{G} \mathbf{G}^H \pmb{\Theta}^H \mathbf{H}))$. Consequently, a good feasible point to the channel capacity is obtained by the optimal phase shift matrix $\pmb{\Theta}^\ast$ of the total path gain maximization as
		\vspace*{-0.1cm}
	\begin{equation}\label{Prob:capacityv1}
			\begin{aligned}
				\pmb{\Theta}^\ast =	\underset{\pmb{\Theta}}{\mathrm{argmax}} \quad & \mathrm{tr}(  \mathbf{H}^H\pmb{\Theta} \mathbf{G} \mathbf{G}^H \pmb{\Theta}^H \mathbf{H}) \\
				\mbox{subject to} \quad& -\pi \leq \theta_k \leq \pi,  \beta_k = 1, \forall k,\\
				&\pmb{\Theta} =  \mathrm{diag}(e^{j\theta_1},\dots,  e^{j\theta_K}).
			\end{aligned}
			\vspace*{-0.1cm}
		\end{equation}
		We observe that the problem~\eqref{Prob:capacityv1} can be solved by the semi-definite relaxation technique \cite{QingQing2019IRS} or
		the alternating direction method of multipliers (ADMM) \cite{Boyu2020} and the diagonal constraint can be relaxed following the steps in \cite{Boyu2020}.
		\subsubsection{Design of the precoding matrix $\mathbf{F}$} For the given solution to the $\pmb{\Theta}$-subproblem, let us define the aggregated channel $\widetilde{\mathbf{H}} = \mathbf{H}^H\pmb{\Theta} \mathbf{G}$ that imposes all the features of the cascaded channels and the phase shifts.  By utilizing the  singular value decomposition, we formulate the SVD of $\widetilde{\mathbf{H}}$ as $\widetilde{\mathbf{H}} = \mathbf{U} \pmb{\Sigma} \mathbf{V}^H$, where $\mathbf{U} \in \mathbb{C}^{N_r \times N_t}$ and $\mathbf{V}  \in \mathbb{C}^{N_t \times N_t}$ satisfy $\mathbf{U}^H \mathbf{U}= \mathbf{I}_{N_t}$ and $\mathbf{V}^H \mathbf{V}= \mathbf{I}_{N_t}$. Besides, $\pmb{\Sigma} = \mathrm{diag}(\lambda_1, \ldots, \lambda_{N_t})$ contains the singular values $\lambda_m, \forall m= 1, \ldots ,N_t,$ with $\lambda_1 \geq \ldots \geq \lambda_{N_t}$.  The optimal solution to the precoding matrix is 
		\vspace*{-0.1cm}
		\begin{equation} \label{eq:Fast}
			\mathbf{F}^\ast = [\mathbf{V}]_{1:N_s} \mathbf{P}^{1/2} = [\mathbf{V}]_{1:N_s}\mathrm{diag}\left(\sqrt{p_1^\ast}, \ldots, \sqrt{p_{N_s}^\ast} \right) ,
			\vspace*{-0.1cm}
		\end{equation}
		where $p_m^\ast$ denotes the optimal fraction of the transmit power assigned to the $m$-th data stream satisfying $\sum_{m=1}^M p^\ast_m = N_s$, and $[\mathbf{A}]_{1:N_s}$ denotes the first $N_s$ columns of matrix $\mathbf{A}$.
		\subsubsection{Design of the equalizer matrix $\mathbf{Z}$} Conditioned on the solutions to the $\mathbf{Z}$- and $\mathbf{F}$- subproblems, the received signal in \eqref{eq:ReceivedSig} can be reformulated by substituting \eqref{eq:Fast} into \eqref{eq:ReceivedSig} and doing some algebraic manipulations as
		\vspace*{-0.1cm}
		\begin{equation} \label{eq:s1}
			\hat{\mathbf{s}} \stackrel{(a)}{=} \sqrt{P/N_s} \mathbf{Z} \mathbf{U} \pmb{\Sigma}\mathbf{P}^{1/2}\mathbf{s}+ \tilde{\mathbf{n}} \stackrel{(b)}{=} \mathbf{s} + \tilde{\mathbf{n}},
		\vspace*{-0.1cm}
		\end{equation}
		where $\tilde{\mathbf{n}} = \mathbf{Z}\mathbf{n}$. In \eqref{eq:s1},  $(a)$ is obtained by the SVD composition of the aggregated channel $\widetilde{\mathbf{H}}$ and the use of optimal precoding matrix $\mathbf{F}^\ast$ in \eqref{eq:Fast}. 
		In order to detect the transmitted signal effective, $(b)$ is obtained by the following solution
		\vspace*{-0.1cm}
		\begin{equation}
			\mathbf{Z}^\ast = ( \pmb{\Sigma}_{N_s}\mathbf{P}^{1/2})^{-1} [\mathbf{U}]_{1:N_s}^H,
			\vspace*{-0.1cm}
		\end{equation}
		where $\pmb{\Sigma}_{N_s} = \mathrm{diag}(\lambda_1, \ldots, \lambda_{N_s})$.
		Notice that the complexity of the algorithm mainly comes from obtaining $\pmb{\Theta}^\ast$ 
		which is in order of $\mathcal{O}(K^3+TK^2)$, where $T$ is the number of iterations needed to reach the convergence from an initial point. Furthermore, the receiver requires the computational complexity of $\mathcal{O}(N_s2^M)$ to decode the modulated signal from $\hat{\mathbf{s}}$. Therefore, the total complexity raised by this communication system is in the order of $\mathcal{O}(K^3+TK^2+N_s2^M)$.
	\vspace*{-0.2cm}
	\begin{remark}
			Even though $\pmb{\Theta}^\ast$, $\mathbf{Z}^\ast$, and $\mathbf{F}^\ast$ are not the optimal solution, they offer an initial mechanism to study the passive and active resource allocation optimization to the networks under smart environment controls. From the equivalence between the sum channel capacity maximization and the minimum mean square error (MMSE) optimization  \cite{van2018large}, the proposed design is expected to attain a low BER as well.\footnote{{ Following the similar steps as in \cite{van2018large}, we can prove that the sum channel capacity maximization and the  mean square error (MSE) minimization share the same globally optimal solution to $\{ \pmb{\Theta}, \mathbf{Z}, \mathbf{F} \}$.}} 
		\vspace*{-0.2cm}
		\end{remark}
	\begin{figure*}[t]
		\centering
		\includegraphics[trim=0cm 0cm 0cm 0cm, clip=true, width=6.0in]{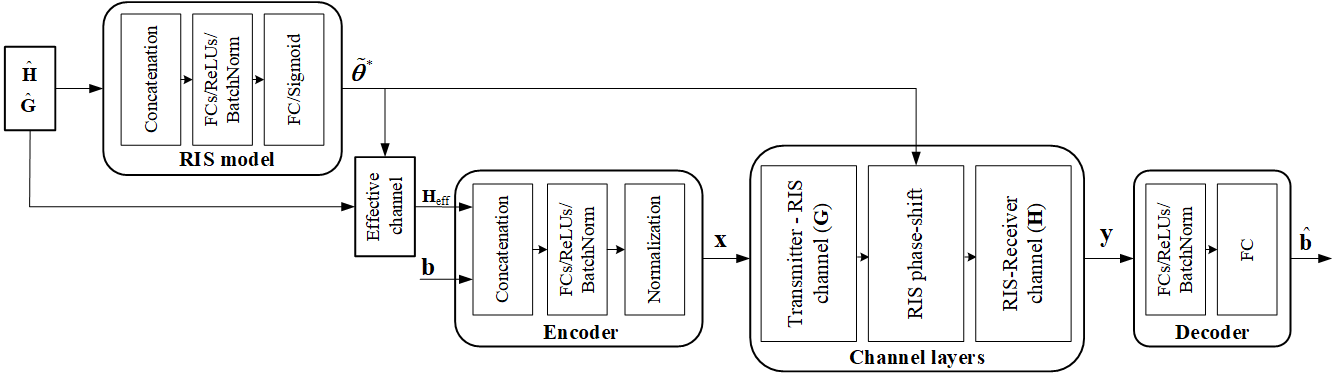}
		\caption{The proposed data-driven framework, where the transmitter, the receiver, and the RIS device are replaced by the neural networks.}
		\label{Fig:AEFramework}
		\vspace*{-0.5cm}
	\end{figure*}
		\vspace*{-0.1cm}
	\section{AUTOENCODER OPTIMIZATION APPROACH} \label{ProposedFramework}
		\vspace*{-0.1cm}
	This section describes the RIS-aided autoencoder-based framework that jointly learns the features of all the active and passive devices as illustrated in Fig.~\ref{Fig:AEFramework}, where the transceiver and the RIS device are replaced by neural networks. 
		\vspace*{-0.1cm}
	\subsection{Preliminary}
		\vspace*{-0.1cm}
	 We consider the design  that minimizes the BER between the decoded signal at the receiver and the transmitted signal at the transmitter. The feasible set of our frame work is characterized by the fact that
	  $0 \leq \|\mathbf{H}^H \pmb{\Phi} \mathbf{G} \mathbf{F} \|_F  \stackrel{(a)}{=} \sqrt{\sum\nolimits_{n=1}^{N_s} \|\mathbf{H}^H \pmb{\Phi} \mathbf{G} \mathbf{f}_n \|_2^2} ,$
	where $\mathbf{f}_n \in \mathbb{C}^{N_t}$ denotes the $n$-th column of matrix $\mathbf{F}$ and $(a)$ is obtained by the Frobenius norm based on the matrix multiplication. Let us denote $\mathbf{o}_m^T \in \mathbb{C}^{N_t}$ is the $m$-th row of matrix $\mathbf{H}^H \pmb{\Phi} \mathbf{G}$, then it holds that
	\vspace*{-0.1cm}
	\begin{equation}
	\begin{split}
	  & \sum\nolimits_{n=1}^{N_s} \|\mathbf{H}^H \pmb{\Phi} \mathbf{G} \mathbf{f}_n \|_2^2 = \sum\nolimits_{n=1}^{N_s} \sum\nolimits_{m=1}^{N_r} \left| \mathbf{o}_m^T \mathbf{f}_n  \right|^2  \stackrel{(a)}{\leq} \\
	   &    \sum\nolimits_{n=1}^{N_s} \sum\nolimits_{m=1}^{N_r} \|\mathbf{o}_m^T \|_2^2 \|\mathbf{f}_n \|_2^2 =  N_s \sum\nolimits_{n=1}^{N_t} \lambda_n \stackrel{(b)}{<} \infty,
	   \end{split}
   	\vspace*{-0.1cm}
	\end{equation}
	 where $(a)$ is obtained by using the  Cauchy–Schwarz inequality and $(b)$ is due to the finite network dimensions and the law of conservation of energy. Consequently, we obtain
	 \vspace*{-0.1cm}
	 \begin{equation}
	 0 \leq \|\mathbf{H}^H \pmb{\Phi} \mathbf{G} \mathbf{F} \|_F \leq  \sqrt{N_s\sum\nolimits_{n=1}^{N_t}\lambda_n}.
	 \vspace*{-0.1cm}
	 \end{equation}
	Besides, the limited power budget at the receiver ensures that $\|\mathbf{Z} \|_F$ is bounded, and therefore the feasible set of our framework is compact. The signal transmission in \eqref{eq:ReceivedSig}  and the signal detection in \eqref{eq:shat} can be expressed as a composition of the continuous mappings. Since  all the requirements of the universal approximation theorem \cite{Goodfellow-et-al-2016} are fulfilled, there exist neural networks to train and predict our considered model.

	 We denote the deep neural networks replaced for the transmitter, receiver, and the RIS device by the encoder, the decoder, and the RIS network, respectively. As shown in Fig.~\ref{Fig:SystemModel} and based on the continuous mappings, the data-driven approach involves the following procedures to train the neural networks in a hierarchical fashion as follows: 
	 \begin{itemize}
	\item[$i)$] The RIS network is a neural network that is trained to predict the desirable phase-shift values based on the channel state information. The detailed interpretation is presented in Sec.~\ref{Sec:RISDesign}.
	\item[$ii)$] The encoder is a neural network that is trained to encode the original data bit stream and predict the transmitted signals (after steering by the beamforming vectors) from both the phase-shift values and the channel information. The detailed interpretation is presented in Sec.~\ref{Sec:TransmitterDesign}.  
	\item[$iii)$] The decoder is a neural network that is trained to  decode the transmitted data bit stream from the received signals. The detailed interpretation is presented in Sec.~\ref{Sec:DecoderDesign}.
	 \end{itemize}
	Conditioned on the fact that the data bit stream are available for the training phase, the design of the data-driven approach is described in detail hereafter.
		\vspace*{-0.1cm}
	\subsection{RIS Network Design} \label{Sec:RISDesign}
		\vspace*{-0.1cm}
	The RIS network is a neural network that is trained to predict the phase shifts gathered in the reflection matrix $\pmb{\Theta}$. We  exploit the channels between the transmitter and the RIS device as well as between the RIS device and  the receiver as the inputs of the neural network\footnote{Since the network training can be performed offline, we assume the availability of CSI. In practice, a classical channel estimation method can be applied to estimate the CSI via the pilot training phase.}. In order to make the framework closer to practical systems, we assume that the perfect instantaneous channels are unavailable. Therefore, the imperfect instantaneous channels $\widehat{\mathbf{G}} \in \mathbb{C}^{K \times N_t}$ and $\widehat{\mathbf{H}} \in \mathbb{C}^{K \times N_r}$ are  defined as follows 
	\vspace*{-0.1cm}
	\begin{equation}
 	\widehat{\mathbf{H}} = \mathbf{H} + \mathbf{H}_{\mathrm{e}} \mbox{ and } \widehat{\mathbf{G}} = \mathbf{G} + \mathbf{G}_{\mathrm{e}},
 	\vspace*{-0.1cm}
	\end{equation}
	where $\mathbf{H}_{\mathrm{e}} \in \mathbb{C}^{ K \times N_r }$ and $\mathbf{G}_{\mathrm{e}} \in \mathbb{C}^{K \times N_t}$ are the corresponding estimation errors, which are assumed to be uncorrelated with $\mathbf{H}$ and $\mathbf{G}$, respectively. The elements of $\mathbf{H}_{\mathrm{e}}$ and $\mathbf{G}_{\mathrm{e}}$ are independent and identically distributed by a circularly symmetric complex Gaussian distribution \cite{Hassibi2003CEtraining} with zero-mean and  variance standing for the channel estimation quality. 
	As shown in Fig.~\ref{Fig:AEFramework}, each realization of the channel estimates $\hat{\mathbf{H}}_r^H$ and  $\hat{\mathbf{G}}$ are first reshaped into a vector of length $2KN_t+2KN_r$ and then fed through a few fully connected layers with rectified linear unit (ReLU) activation functions. To prevent overfitting problems and {enable} efficient training \cite{Sergey2015BatchNorm}, a batch normalization layer is inserted between each pair  of the fully connected layers. 
	The predicted phase shift vector is given as $\tilde{\pmb{\theta}}^{\ast} = \{ \tilde{\theta}_1^{\ast},\tilde{\theta}_2^{\ast},...,\tilde{\theta}_K^{\ast} \}$, followed by the predicted reflection matrix $\widetilde{\pmb{\Theta}}^\ast = \mathrm{diag}( e^{j\tilde{\theta}_1^\ast},\dots, e^{j\tilde{\theta}_K^\ast}) $.
	We notice that the parameter settings for the RIS network are given in Table~\ref{AE-para}. Furthermore, the predicted reflection matrix $\widetilde{\pmb{\Theta}}^\ast$ is then utilized to  formulate the estimated cascaded channel of the indirect link channel from the transmitter to the receiver through the RIS device as $\mathbf{H}_{\mathrm{eff}} = \widehat{\mathbf{H}}^H \widetilde{\pmb{\Theta}}^\ast \widehat{\mathbf{G}}$.
	\begin{table}[t] 
		\centering
		\caption{Parameter setting for the encoder, decoder, and RIS model.}
		\begin{tabular}{*5l}
			\hline
			Layers & Encoder & RIS model & Decoder\\
			\hline
			Input & \thead{$MN_s+$  $2N_tN_r$} & $2KN_t+2KN_r$ &$2N_r$\\
		$1$st fully con. layer & 1024 & 256 & 512 \\
			BatchNorm1d & 1024 & 256 & 512\\
			Activation function & ReLU & ReLU & ReLU \\
			$2$nd fully con. layer & 1024 & 256 & 512 \\
			BatchNorm1d & 1024 & 256 & 512\\
			Activation function & ReLU & ReLU & ReLU \\
			$3$rd fully con. layer & $2N_t$ & 256 & $MN_s$ \\
			BatchNorm1d & - &256 &-\\
			Activation function &-&ReLU&-\\
			$4$th fully con. layer & - & $K$ & - \\
			Activation function &-&Sigmoid&- \\
			Normalization & $2N_t$ & - & - \\
			\hline
		\end{tabular} \label{AE-para}
	\end{table}

   	\vspace*{-0.1cm}
	\subsection{Transmitter Design} \label{Sec:TransmitterDesign}
		\vspace*{-0.1cm}
	In our transmitter design based on a neural network, the estimated cascaded channel $\mathbf{H}_{\mathrm{eff}}$ is considered as the input of the autoencoder along with the data bit stream. More specifically, as shown in Fig.~\ref{Fig:AEFramework}, the estimated cascaded channel is concatenated with the data bit stream $\mathbf{b}$ and fed to the autoencoder. By this mechanism, the autoencoder can inherit the channel state information to combat fading and greatly improve the system performance. In our framework, the data bit stream $\mathbf{b}$ is divided into  the $N_t$ streams of one-hot vector, each representing one of the $M$ possible modulated data signals. The input is then fed through multiple fully connected layers with the ReLU activation function and batch-normalization layers. Note that the last fully connected layer has a size of $2N_t$ corresponding to the real and imaginary part of the modulated data symbols at the transmitted antennas. The output of the auto    encoder is then reshaped to generate the complex transmitted signal vector $\mathbf{x} \in \mathbb{C}^{N_t}$. Before transmitting the signal, we apply an average power constraint by using a normalization layer, which is a custom layer and can be considered as a neural layer without any trainable parameters. The output of the normalization layer is 
	\vspace*{-0.1cm}
	\begin{equation} \label{eq:x}
		\mathbf{x} = P B^{1/2}\Big(\sum\nolimits_{i=1}^{B}\|\mathbf{x}'_i \|_2^2\Big)^{-1/2} \mathbf{x}', 
		\vspace*{-0.1cm}
	\end{equation}
	where $B$ is the mini-batch size and $\mathbf{x}' \in \mathbb{C}^{N_t}$ is the output of the last fully connected layer. In \eqref{eq:x}, the predicted signal $\mathbf{x}$ is the transmitted signal with a  transmit power level.
	\vspace*{-0.1cm}
	\subsection{Channel Layers}
	\vspace*{-0.1cm}
	In order for the system to build an end-to-end framework, we design several custom layers to simulate the data propagation via the propagation channels with the presence of the RIS device as shown in Fig.~\ref{Fig:AEFramework}. Similar to the normalization layer of the transmitter, these channel layers are custom layers with untrainable parameters to perform the complex multiplication between signals and channels. Different from \cite{Tugba2021AE-RIS}, in the training process, the channels are changing along with every transmitted symbol. {Therefore, the trained neural networks can encode the data bit streams with the aware of channel condition instead of only a function of the bit information as in the conventional modulation schemes.}
		\vspace*{-0.1cm}
	\subsection{Decoder Design} \label{Sec:DecoderDesign}
		\vspace*{-0.1cm}
	The decoder is a fully connected neural network whose input is the received signal $\mathbf{y} \in \mathbb{C}^{N_t}$. The input data in the complex field are first stacked into a vector including both the real and imaginary parts.
	After that, the stacked data go through the multiple fully connected layers with the ReLU activation function and the batch normalization layer between each pair of two fully connected layers. The output of the neural network responsible for the decoder is separated to form the recovered data  $\hat{\mathbf{b}} = [\hat{\mathbf{b}}_1,\cdots,\hat{\mathbf{b}}_{N_t}]$ of the original one-hot data $\mathbf{b} = [\mathbf{b}_1,\cdots,\mathbf{b}_{N_t}]$. We stress that there are total $M$ output classes in each data stream by separating the decoded signal. Furthermore, the equalization step is done at the receiver without any channel knowledge. In more details,  Table~\ref{AE-para} shows the parameter setting of the decoder in detail.
	\vspace*{-0.1cm}
	\subsection{Optimization Process}
	\vspace*{-0.1cm}
    
As the main theme of an autoencoder, we jointly learn the parameterized encoder, decoder, and RIS device by minimizing the loss function for a given modulation scheme as
\vspace*{-0.1cm}
\begin{equation} \label{Pro:Loss}
\begin{aligned}
& \underset{\{f,g,r\}·}{\mathrm{minimize}} && \mathcal{L}_{\mathrm{AE}} (\psi_f,\psi_g,\psi_r),\\
& \mbox{subject to} && \mathbf{s} \in \mathcal{M}, 
\end{aligned}
\vspace*{-0.1cm}
\end{equation}
where $\mathcal{M}$ is the finite constellation set defined by the M-QAM in this paper. $\psi_f, \psi_g,$ and $\psi_r$ are the parameters of the encoder $f$, the decoder $g$, and the RIS network $r$, respectively. {Since the data bits are represented by one-hot vectors, the detection of these bits can be regarded as a typical classification problem. Therefore, cross-entropy loss is readily used for the network optimization.} The loss function $\mathcal{L}_{\mathrm{AE}} (\psi_f,\psi_g,\psi_r)$ is formulated as
\vspace*{-0.1cm}
	\begin{equation} \label{eq:LossFunc}
	\mathcal{L}_{\mathrm{AE}} (\psi_f,\psi_g,\psi_r) = \sum\nolimits_{i=1}^{N_s}\alpha_i\mathcal{L}_i(\psi_f,\psi_g,\psi_r),
\vspace*{-0.1cm}
	\end{equation}
where $\mathcal{L}_i(\psi_f,\psi_g,\psi_r)$ is  the loss function for the $i$-th  data stream, which is defined as
	\begin{multline} \label{eq:Lossi}
    		\mathcal{L}_i(\psi_f,\psi_g,\psi_r) = \\  -\frac{1}{B}\sum\nolimits_{m=1}^{B}\sum\nolimits_{n=0}^{M-1}\mathrm{log}\left(\frac{\mathrm{exp}\left([\mathbf{b}_i]_n \right)}{\mathrm{exp}\left(\sum\nolimits_{j=0}^{M-1}[\mathbf{b}_i]_j \right)}\right)p([\hat{\mathbf{b}}_i]_n),
	\end{multline}
where $[\mathbf{b}_i]_n$ denotes the $n$-th bit of the one-hot data $[\mathbf{b}_i]$, $p([\hat{\mathbf{b}}_i]_n)$ is the output of the last fully connected layer in the decoder which can be regarded as the probability of the $(n+1)$-th possible modulated signal for data $[\mathbf{b}_i]$,  and $\mathrm{exp}(\cdot)$ is the exponential function. In \eqref{eq:LossFunc},  $\alpha_i \geq 0$ is a weight associated with the $i$-th loss function and satisfied $\sum_{i=1}^{N_s} \alpha_i =1$. Notice that an equal weight setting, i.e., $\alpha_i = 1/N_s, \forall i,$ may lead to the unfair performance between each stream. To deal with unfair performance issues, we apply a dynamic scheme where the weights for the loss function are updated in the $t$-th mini-batch as follows
\begin{equation}
		\alpha_i^t = \mathcal{L}_i^t(\psi_f,\psi_g,\psi_r)/\mathcal{L}^t_{AE}(\psi_f,\psi_g,\psi_r), \quad \forall t.
\end{equation}

In this context, the autoencoder would be trained to obtain a balanced loss among the data streams. We exploit the stochastic gradient descent to train the autoencoder. Hence, the weights and biases are updated based on solving \eqref{Pro:Loss} and through the back propagation. 

	\begin{remark}
	The complexity of  fully connected neural networks  grows with the size of input and output. Therefore, the complexity of the autoencoder is in the order of $\mathcal{O}(MN_s + N_tN_r + K(N_t+N_r))$ which comprises of encoder, decoder and RIS model. From these calculations,  our model obtains much lower complexity compared with the model-based algorithm.
	\end{remark}
	
	\begin{figure*}[t]
	\begin{minipage}{0.33\textwidth}
		\centering
		\includegraphics[trim=2cm 0.3cm 2cm 0.5cm, clip=true, width=2.75in]{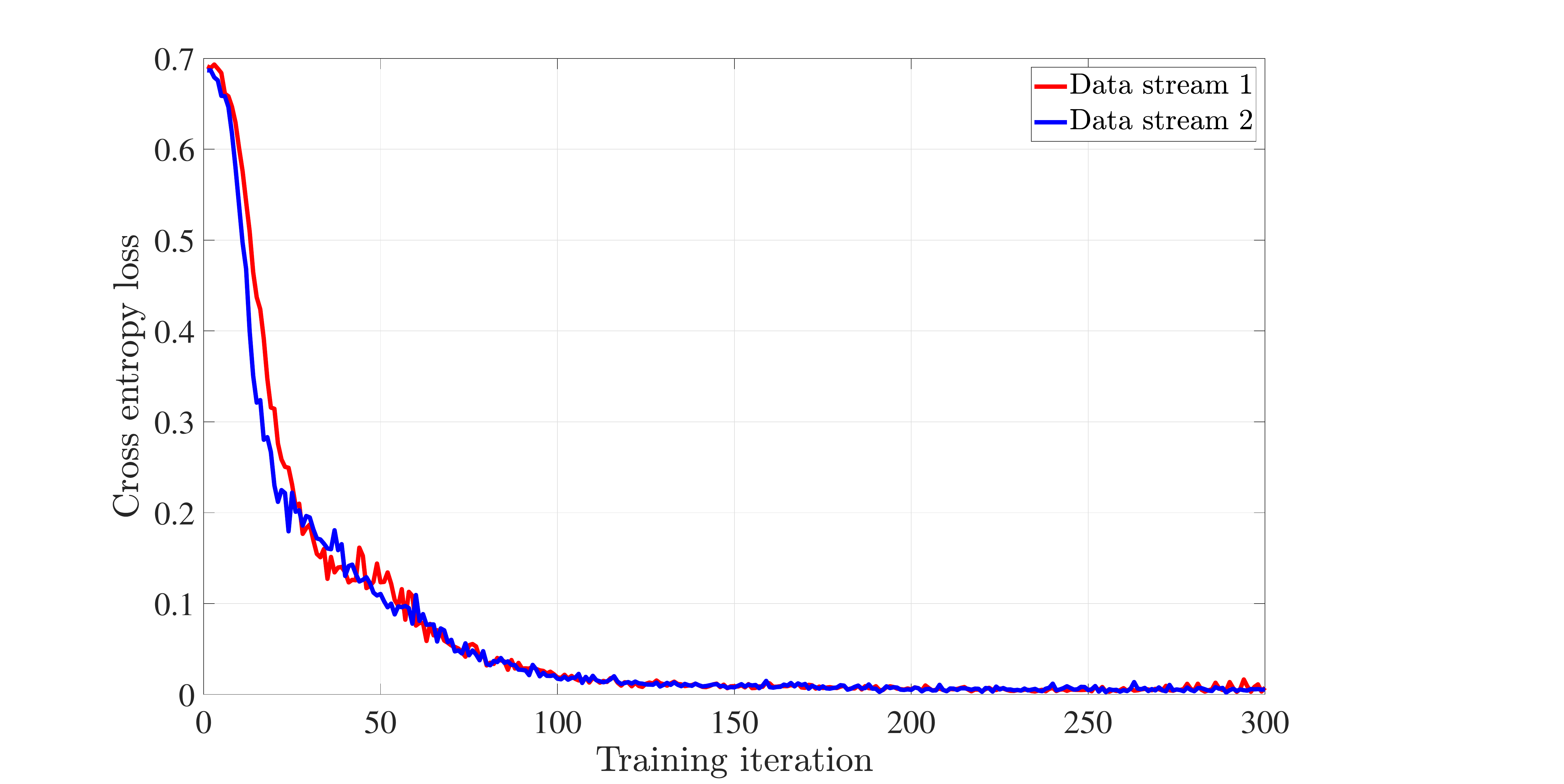} \vspace*{-0.4cm}\\
		$(a)$
		
	\end{minipage}
	\begin{minipage}{0.33\textwidth}
		\centering
		\includegraphics[trim=2cm 0.3cm 1.9cm 0.5cm, clip=true, width=2.75in]{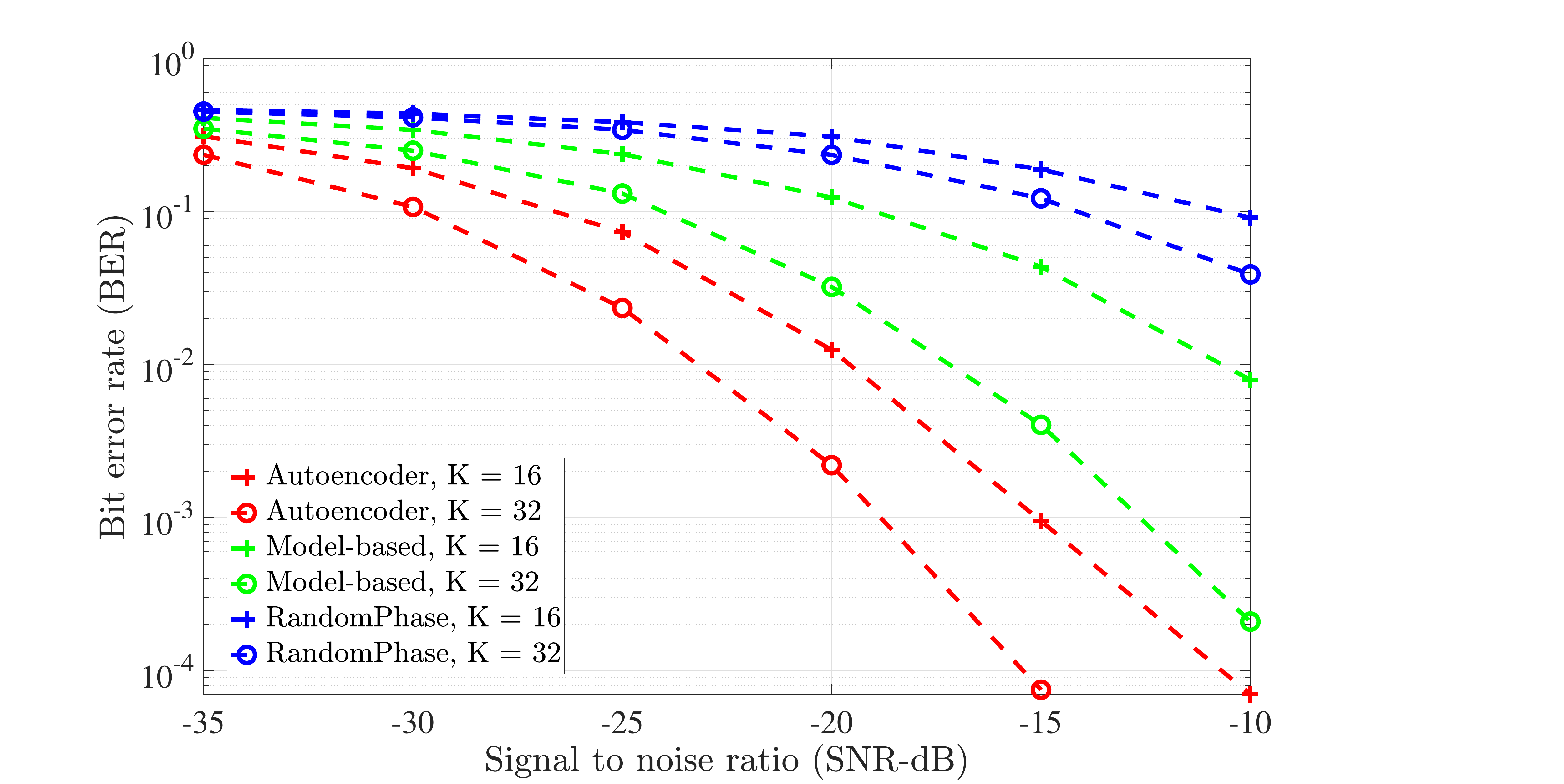} \vspace*{-0.4cm}\\
		$(b)$
	\end{minipage}
	\begin{minipage}{0.33\textwidth}
		\centering
		\includegraphics[trim=2cm 0.3cm 2cm 0.5cm, clip=true, width=2.75in]{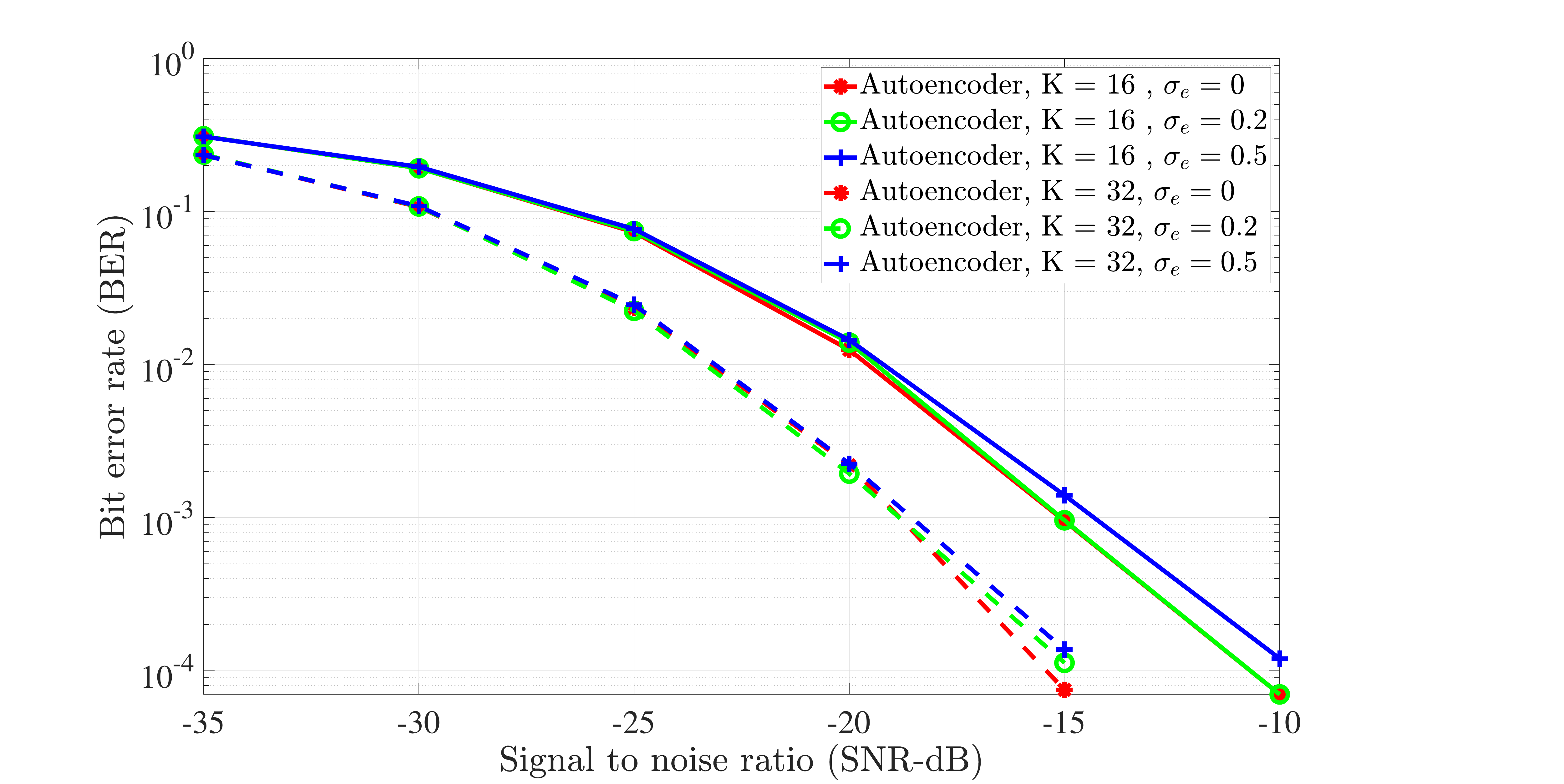} \vspace*{-0.4cm} \\
		$(c)$
	\end{minipage}
	
	\caption{The system performance: $(a)$ The cross entropy versus the training iteration; $(b)$ The BER performance as a function of the SNR value for both the model-based and autoencoder approaches; and $(c)$ The BER performance of the autoencoder approach under imperfect channel state information.}
	\vspace*{-0.6cm}
	\label{Fig:Result}
	\end{figure*}
	\vspace*{-0.1cm}
	\section{Numerical Results} \label{Simulation}
	\vspace*{-0.1cm}
    \begin{table}
	    \centering
	    \caption{Average Running Time of the model-based and Autoencoder Approaches}
    	\begin{tabular}{ | m{2.1cm} | m{1.6cm}| m{1.6cm} | m{1.5cm}|} 
          \hline
          \backslashbox{$K$}{Models}& Model-based  & Autoencoder \\ 
           
          \hline
          16 & 3.480 ms & 0.251 ms\\ 
          \hline
          32 & 7.131 ms & 0.255 ms\\
          \hline
        \end{tabular} \label{Running-Time}
    \end{table}
	We  evaluate the performance of our considered frameworks by a  $4\times 2$ MIMO system transmitting $N_s = 2$ data streams with equal transmitted power on each data stream $P = 4$~[W]. The RIS device is equipped with eight phase-shift elements. The binary phase-shift keying (BPSK) modulation/demodulation are used at the transmitter and the receiver, respectively. 
	In order for the system to train our deep neural networks,  $200000$ different data symbols along with $200000$ channel realizations with $\sigma_e = 0.1$ are used as data for the training phase. We define SNR as the ratio between the transmit signal power and the noise power. The SNR is fixed at $5 \mathrm{dB}$, while it will be varied in the testing phase. The Adam optimizer is selected to train the neural networks. The hyper-parameters are chosen as: the number of epochs is $10$, the mini-bath size is $1000$, and the learning rate is $0.0002$. Conditioned on the phase-shift design, the following benchmarks are considered for comparison:
	 $i)$ \textit{RIS-assisted Joint Transmitter and Receiver Design} is the model-based  approach  presented in Sec.~\ref{Sys-Model} and it is denoted as ``Model-based" in the figures;
	    $ii)$ \textit{RIS-assisted Autoencoder} is the data-driven approach presented in Sec.~\ref{ProposedFramework} and it is denoted as ``Autoencoder" in the figures.
In Fig.~\ref{Fig:Result}$a$, the cross entropy losses for all data streams are plotted as the function of the training iteration. As can be seen, loss functions of both data streams converge after a hundred training iterations. Thanks to the adaptive weight applied in loss function, the losses for both data streams are close to each other in every iteration. To evaluate our proposed model, in Fig.~\ref{Fig:Result}$b$, we plot the BER as a function of the SNR for all the considered benchmarks with the two different number of phase shifts. In addition, we also plot the performance of ``Model-based" approach with random phase-shifts as a performance bound. As illustrated, thanks to the cooperation between transmitter, receiver and RIS, ``Autoencoder" yields the lowest bit error rate in all SNR values. Moreover, when the number of RIS reflecting elements is increased, the performance of both models are greatly improved. This is very intuitive since ``Model-based" can achieve higher channel capacity with more RIS reflecting elements. Even though higher capacity does not lead to optimal bit error rate, as mentioned in Remark 1, higher capacity is expected to attain a higher BER performance. For our model, given the increase in the number of phase-shifts,  the end-to-end framework can encode and decode data in a more flexible way to combat noise effect and reduce error rate in decoding data. Additionally, as listed in Table.~\ref{Running-Time}, our model is dozens of times faster than model-based approaches with 200 iterations which shows its complexity efficiency.

To illustrate the robustness of our model to imperfect CSI, we  test the proposed model with different channel error variance values with $K = 16, 32$ as in  Fig.~\ref{Fig:Result}$c$. Surprisingly, in both setups the performance of ``Autoencoder" remains almost unchanged with different channel error variance. These results show that ``Autoencoder" is very robust to imperfect CSI. {This robustness comes from the fact that ``Autoencoder" is trained with imperfect CSI. Moreover, since various channel realizations are utilized in training process, the encoder can learn to encode data in a way that is robust to various channel conditions.} Hence, the effect of imperfect CSI can be reduced. From the results, we can conclude that our proposed model can guarantee a good performance when the perfect CSI is not available at the transmitter.

\vspace*{-0.2cm}
\section{Conclusion}
\label{conclusions}
\vspace*{-0.2cm}
In this paper, an autoencoder approach for RIS-based MIMO system has been presented to enhance to bit error rate performance of the system. We replaced the transceiver architecture and the RIS model with three FCNN models and trained them jointly with the objective of minimizing the BER of estimated signal at the receiver. By utilizing the BPSK modulation scheme for two independent data streams, the performance of the proposed framework has been compared with the conventional RIS designs in terms of bit error rate performance. Due to the cooperation between the transmitter, receiver and the RIS operation, our framework showed the superior improvements in detecting signals at the receiver. 
	
\vspace*{-0.0cm}	
	
	\bibliographystyle{IEEEtran}
	\bibliography{refs}


\end{document}